\documentclass[aps,prl,twocolumn,showpacs,superscriptaddress]{revtex4-2}
\usepackage{graphicx}  
\usepackage{dcolumn}   
\usepackage{bm}        
\usepackage{amssymb}   
\usepackage{float}
\usepackage{amsmath} 
\usepackage{xcolor}
\usepackage{ulem} 
\usepackage{xr}
\usepackage{comment}

\usepackage{soul}

\begin{document}

\title{Morphogenesis of cheese flowers through scraping}
\author{J.~Zhang}
\email{jishen.zhang@espci.fr}
\affiliation{PMMH, CNRS, ESPCI Paris, Université PSL, Sorbonne Université, Université de Paris, F-75005, Paris, France}
\author{A.~Ibarra}
\affiliation{PMMH, CNRS, ESPCI Paris, Université PSL, Sorbonne Université, Université de Paris, F-75005, Paris, France}
\author{B.~Roman}
\affiliation{PMMH, CNRS, ESPCI Paris, Université PSL, Sorbonne Université, Université de Paris, F-75005, Paris, France}
\author{M.~Ciccotti}
\email{matteo.ciccotti@espci.fr}
\affiliation{SIMM, CNRS, ESPCI Paris, Université PSL, Sorbonne Université, F-75005, Paris, France}
\date{\today}

\begin{abstract}

The “Tête de moine” Swiss cheese is generally served by scraping the surface of a cylindrical loaf with a sharp tool. This produces thin sheets of cheese that are strongly wrinkled at the edge, resembling frilly flowers and enhancing the tasting experience. In this work we unveil the physical mechanisms at play in this scraping-induced morphogenesis. We measure the deformation of the cheese during scraping  and show that plastic deformation occurs everywhere, but find a larger plastic contraction in the inner part of the flower, causing its buckling into shape. We show that it surprisingly derives from the lower friction coefficient evidenced on the cheese close to its crust. Our analysis provides the tools for a better control of chip morphogenesis through plasticity in the shaping of other delicacies, but also in metal cutting.

\end{abstract}

\maketitle

\begin{figure}[ht]
  \includegraphics[width=8cm]{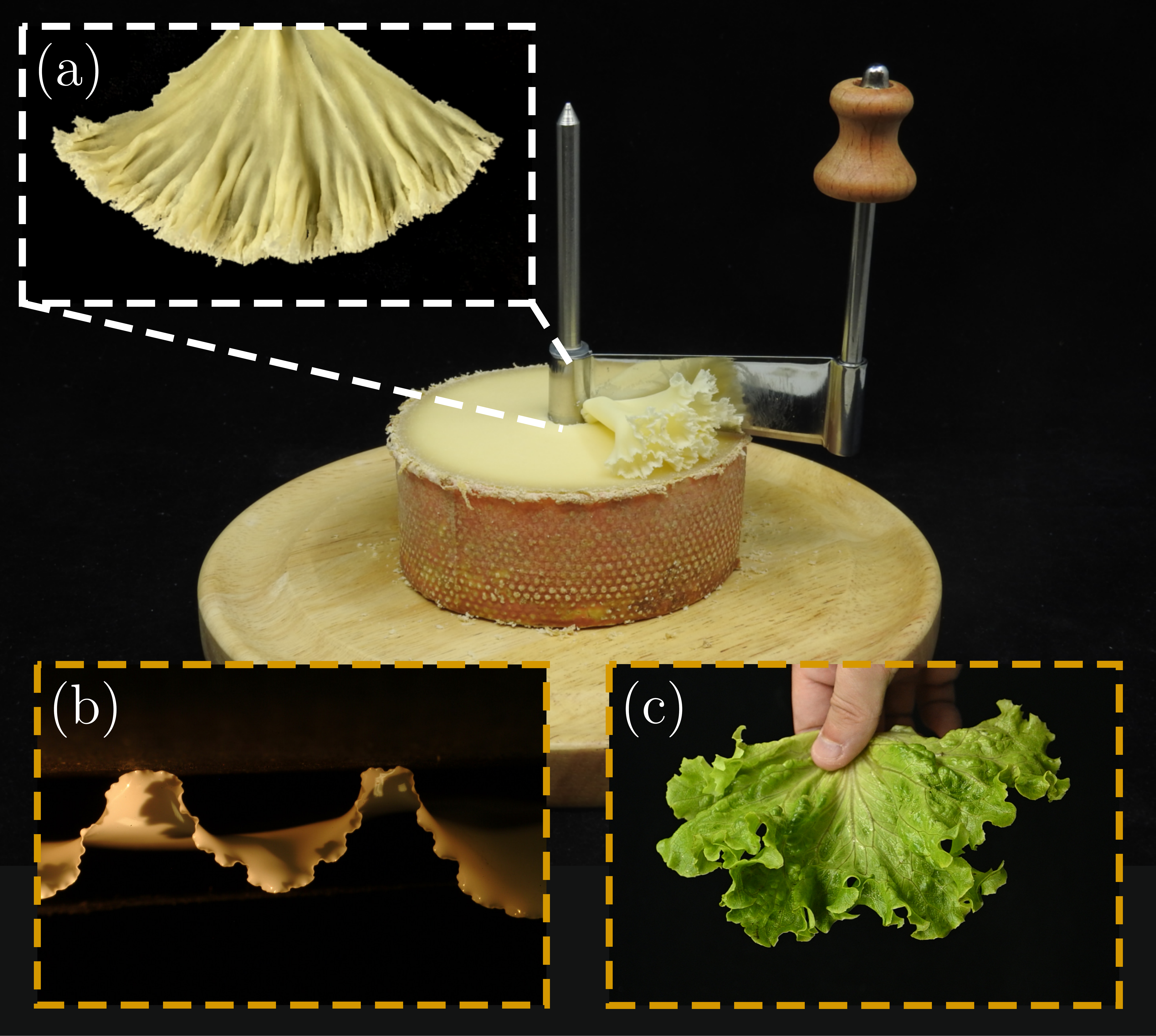}
  \caption{\label{fig:generalFig} (a) Layer of the cheese \textit{Tête de moine} with wrinkly edge, after being scraped using the cheese slicer ``\textit{la Girolle}". (b): Edge wrinkling of a torn plastic sheet; (c) Wavy edge of a Blue Star Fern leaf.}
\end{figure}

-- \textit{Introduction.} The way to serve cheeses differs greatly depending on the mechanical property of each type yet most employed techniques consist of spreading and slicing. The high softness of Blue cheese, mature Camembert, Brie or Époisses allows an effortless spread on bread, while cheeses with higher hardness such as Cheddar, Romano, Comté or Parmesan are served in slice or dice. The Monk's head (Tête de moine) is a cylindrical cheese block originated from the Swiss Jura mountains, often scraped with a slicer called \textit{la Girolle}. The rotating blade scrapes a thin layer of the cheese from the top into a pack of wrinkly layers~(Fig.~\ref{fig:generalFig}),
which is 
considered not only to be aesthetically unique, but also crucial to the culinary experience: the high surface-to-volume ratio of the ethereal cheese packs enhances the release of aromatic flavors and provides a tender mouthfeel.

Similar frilly edges with multiple superimposed wavelength are observed in leaves (Fig.~\ref{fig:generalFig}c), fungi, corals and some marine invertebrate, but also in torn plastic sheets~\cite{sharon2002buckling} (Fig.~\ref{fig:generalFig}b) and were interpreted as a generic motif appearing in non-Euclidean elastic surfaces with strong negative Gaussian curvature~\cite{Yamamoto21}. Such complex surfaces can be produced by a rather simple but non-flat metric~{\cite{sharon2002buckling,Klein07}, as the distribution of distances along a surface dictates its Gaussian curvature~\cite{struik}. For example, plant petals acquire a complex shape because differential in-plane growth \cite{Rolland-Lagan03} produces incompatible length distribution (also called non-Euclidian metrics~{\cite{Efrati07}). A different mechanism leading to similar shapes occurs during the tearing of thin plastic sheets, where irreversible plastic stretching increases the length of lines parallel to the boundary~\cite{sharon2002buckling} strongly when closer to the free edge. Do these ``cheese flowers" fall into the same category of morphogenesis ? What physical mechanism is responsible for producing this shape ?

In this letter, we first provide a thorough characterization of the metric changes in the cheese flowers obtained under well controlled cutting conditions. We then consider the mechanism of plastic scraping and show how the inhomogeneous metric results from the variation of mechanical properties along the radius of the cheese loaf. 

-- \textit{Experimental setup.}
 In the original cutting tool~\textit{la Girolle} (Fig.~\ref{fig:generalFig} top), the cheese is skewered on a steel cylindrical rod fixed in the center of a wooden base-plate. The blade fixture is mounted on the rod in order to keep the cutting edge horizontal, while being free to move vertically during the cutting under the action of an applied weight. The cutting edge is machined to have an overhanging (negative) rake angle $\alpha=-14.7^{\circ}$ with respect to flat cheese surface (cf.~Fig.~\ref{fig:merchant_snapshot}b). Our experiment is set to achieve a steady-state imaging of the chip cutting process (cf.~Fig.~\ref{fig:initialDepth_setup}): the blade fixture is geared at a fixed position and the base is motorized to rotate at a constant speed $\Omega=1.14$ rad/s. The Monk's head cheese wheels used in the present work were selected from a single brand (Fromagerie de Bellelay) and bought from a single distributor, with a selected refining age between three and six month. The initial cylindrical shape has a typical radius of 50 mm and height of 120 mm. Each cheese is initially cut in two halves in order to work in the core region. The vertical cutting force is set by adding variable masses on a plate mounted on top of the blade (cf.~Fig.~\ref{fig:initialDepth_setup}).

\textit{Depth of cut.} Fig.~\ref{fig:initialDepth_setup} (inset) reports the measured depth of cut as a function of the total applied weight, normalized by the length of the blade contact on the cheese. The depth of cut $h_0$ was estimated by counting the number of full spins required to cut a given depth of one millimeter in the cheese. We can remark that there is no appreciable cutting below a threshold force of 0.16 N/mm, and then the depth of cutting is linearly increasing with the applied load. For the following measurements, we selected the applied weight to fix a depth of cut of $h_0 = 0.32$ mm in order to obtain stable cheese flowers that can be conveniently manipulated. 

\begin{figure}[ht]
\includegraphics[width=8cm]{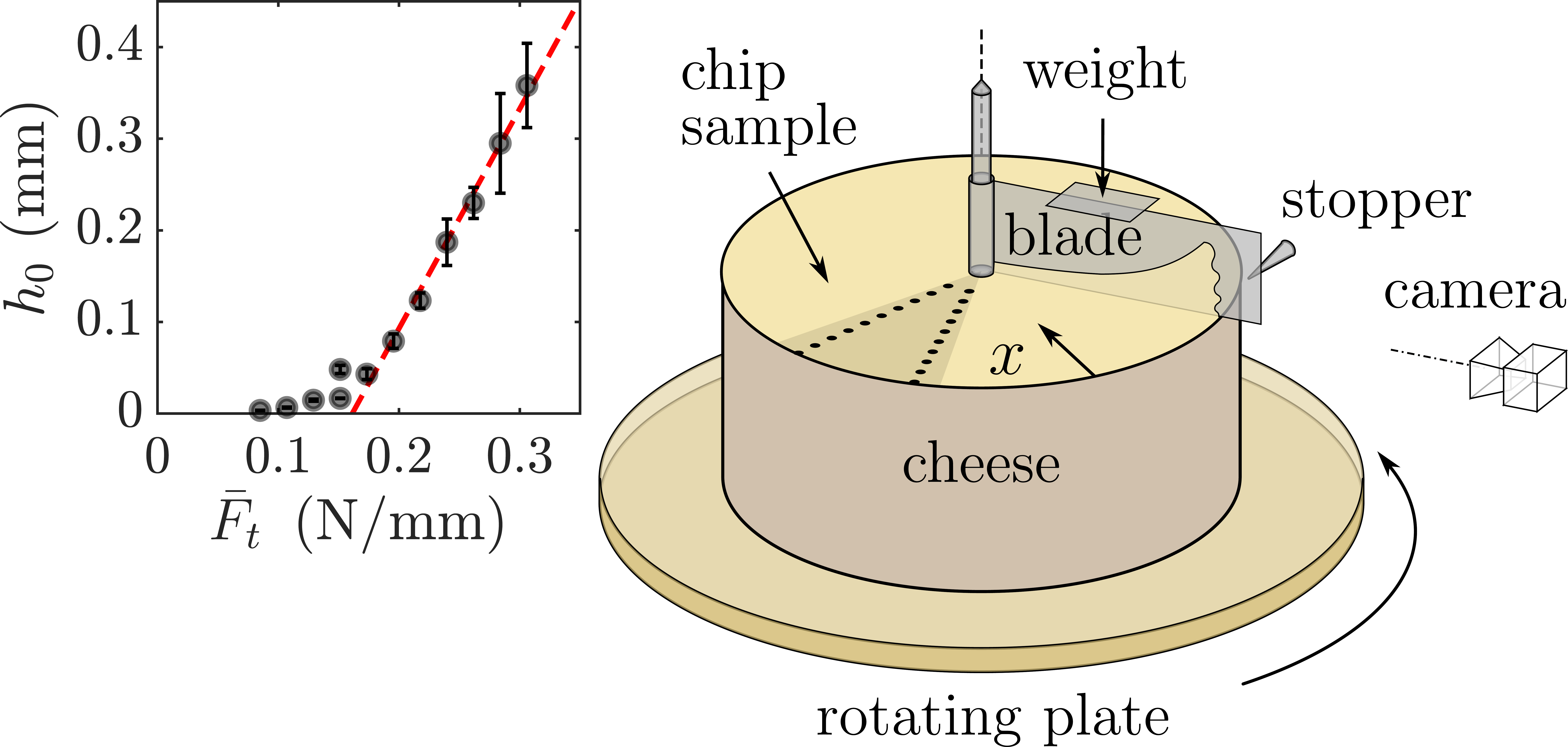}
\caption{\label{fig:initialDepth_setup} Schematic illustration of the experimental setup. Inset:  depth of cut $h_0$ as a function of vertical load per unit length of the blade $\overline{F}_t$.}
\end{figure}

\begin{figure}[ht]
\centering
\includegraphics[width=7cm]{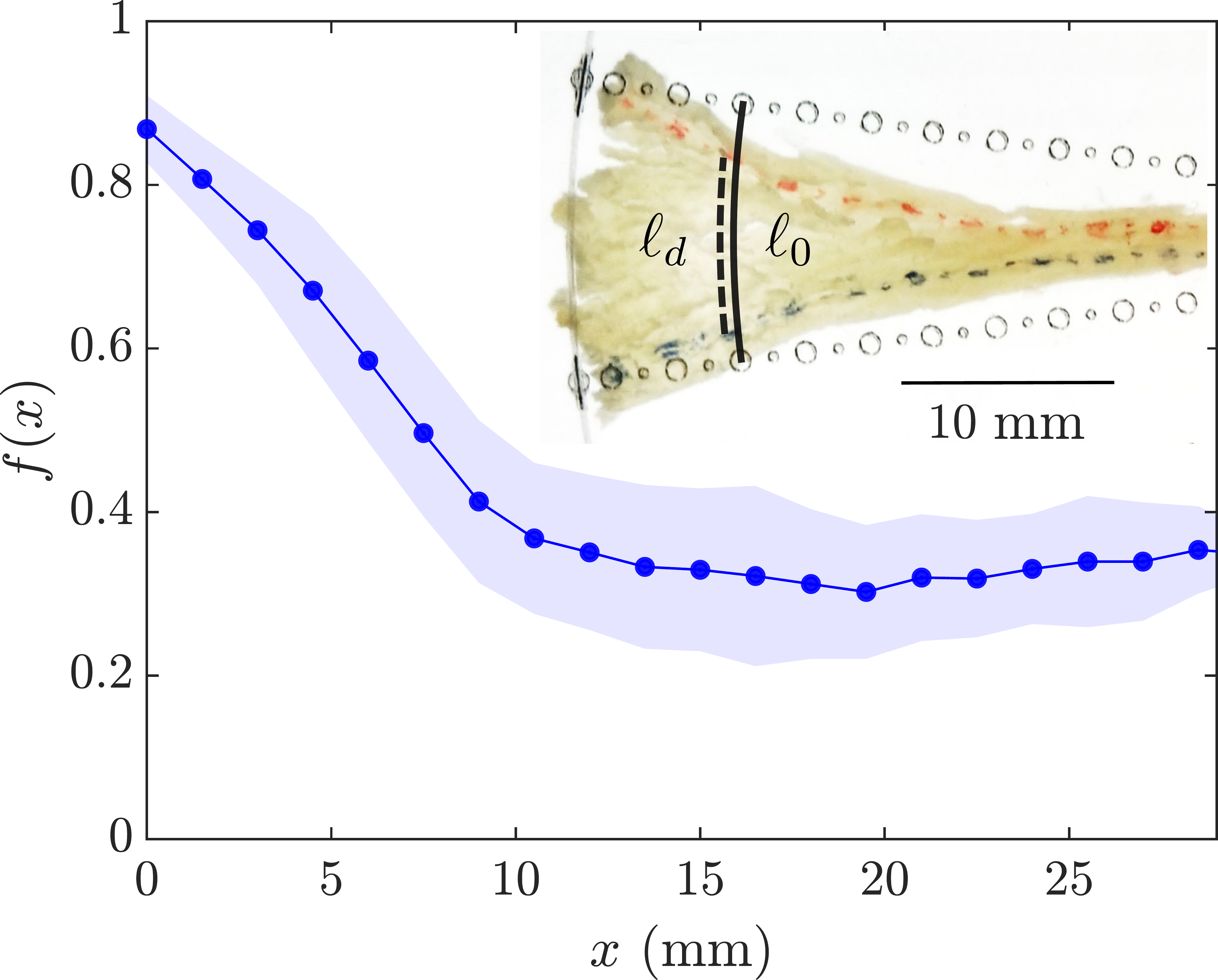}
\caption{\label{fig:CR} Picture of a flattened cheese flower chip (inset) and associated metrics. The difference between the two color deformed dotted lines (red and blue) and their original position along straight lines (black circles) is used for the measurement of the metric change $f(x)=\ell_d/\ell_0$ versus the radial distance $x$ from the edge periphery. The blue dots and line in the main figure represent the local average, and the shaded area the standard deviation.}
\end{figure}

-- \textit{Radial metric measurements.}
Due to the rotational symmetry, we expect the metric to depend only on the radial distance to the center. Radial metrics variations in the scrapped chip samples were measured as follows. We first polish the surface of the cheese by cutting with a low weight close to the threshold. We then draw two straight radial dotted lines forming a small angle of $ 15^\circ$ on the target part of the chip to be measured. We start the cutting remotely upstream in order to reach the steady state depth before the target chip. Then we trim the target chip along the dotted lines and flatten it gently as shown in Fig.~\ref{fig:CR} inset.   
Tracking the displacement of the dots we can observe that each arc $\ell_0$ is deformed into an arc $\ell_d$ which is shortened by a factor $f=\ell_d/\ell_0$, while the radial displacement is very small and will be neglected. The metric is thus defined by the single function $f(x)$ 
of the radial distance $x$ from the periphery, as plotted in Fig.~\ref{fig:CR}. 
Close to the periphery the metric is hardly distorted ($f(0)\sim 0.9$ is close to $1$), but the ratio $f(x)$ decreases monotonically over a distance of 10 mm from the periphery, before reaching a fairly constant plateau region evidencing a large contraction by a factor almost 3 ($f \sim 0.35$). This distance was observed to be independent of the cheese radius, and it soundly corresponds to an external boundary layer of the cheese which is affected by the drying process \cite{prentice1993cheese, gunasekaran2002cheese}. The mismatch of longitudinal length between the periphery and the core material is somewhat reminiscent of the metric distortion in torn plastic sheets \cite{sharon2002buckling}.  But in that case, the mismatch is caused by an increased plastic stretch of the periphery ($f(0) $ typically of order 2, with a plateau $f=1$), whereas for the cheese flowers, it is due to an increased plastic contraction in the inner region. 
Nevertheless, we conclude that the complex buckled shape of the cheese flower results here from the same type of internal geometrical frustration \cite{sharon2002buckling,siefert2019bio}: the buckle emergence is intrinsically associated to a spatial metric variation. If the metric change $f$ was constant everywhere, the ribbon would have zero Gaussian curvature and would stay flat. 

-- \textit{Chip cutting mechanism}. 
In the following we wish to elucidate the cause of inhomogeneous shrinking of the cheese ribbon.
A natural idea is to imagine that the cutting mechanism changes from elastic close to the harder periphery to plastic in the soft core of the cheese.
\begin{figure}[ht]
\includegraphics[width=7.8cm]{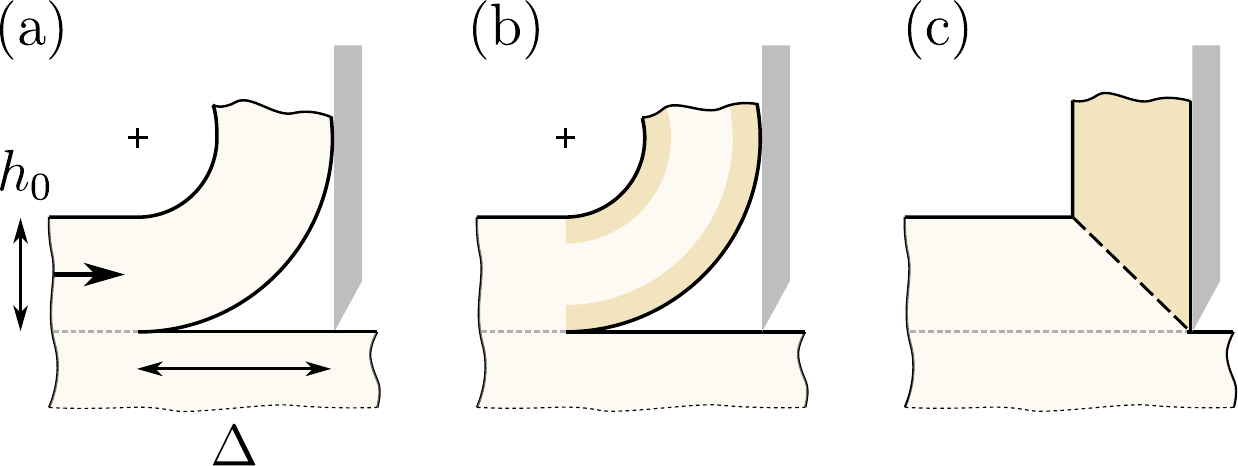}
\caption{\label{fig:sketchTransition} Sketch of elasto-plastic chip deformations by cutting. The blade is set vertical to the sample surface for simplification. (a) Elastic bending. (b) Elastic-plastic bending. (c) Plastic shearing. Plastic zones are highlighted with dark color.}
\end{figure}
To address this issue, we can benefit from an extensive literature about the complex chip shapes and deformation mechanisms in the metal cutting community \cite{trent2000metal,shaw2005metal,stephenson2018metal,lechenault2023soft}, and report the essential physical arguments here. 
Cutting a layer into an elastic material results in cleavage, i.e.\ the elastic bending of the chip with a crack at a distance $\Delta$ ahead of the blade tip as in Fig.~\ref{fig:sketchTransition}a, where a vertical rake is considered for simplicity, without loss of generality. The chip curls up with a typical curvature $1/\Delta$. The cleavage distance $\Delta$ can be found  by balancing fracture energy  $\Gamma \Delta$ and bending energy $ Eh_0^3\Delta/\Delta^2$\cite{obreimoff1930splitting}, so that $\Delta \sim \sqrt{Eh_0^3/\Gamma}$, where $E$ is Young's modulus.
This elastic scenario breaks down if plastic yielding is reached in bending (cf.~Fig.~\ref{fig:sketchTransition}b), when the maximum stress in the chip $E h_0/\Delta$ attains the yield stress of the material $\sigma_y$, i.e.\ when:
\begin{equation}
\frac{h_0 \sigma_y^2}{ E \Gamma} = \frac{h_0}{\ell_D} \lesssim 1,
\label{eq:Dugdale}
\end{equation}
where $\ell_D = E\Gamma/\sigma_y^2$ is known as the Dugdale length \cite{dugdale1960yielding}, which sets the physical length-scale 
for the balance of fracture and plastic processes and represents the distance to the crack below which plastic yielding dominates. 
A precise calculation \cite{williams2016fundamentals} 
shows that the elastic cleavage takes place when 
$h_0 / \ell_D  > 6$, and plastic bending occurs below this value. 
But when the depth of cut $h_0$ becomes comparable to the Dugdale length ($h_0 / \ell_D \leq 2$) a different yielding mode takes place 
leading to a homogeneous plastic shearing through the whole thickness of the chip, as sketched in Fig.~\ref{fig:sketchTransition}c, in agreement with experiments on several materials \cite{aghababaei2021cutting}.

\begin{figure}[ht]
\includegraphics[width=8.5cm]{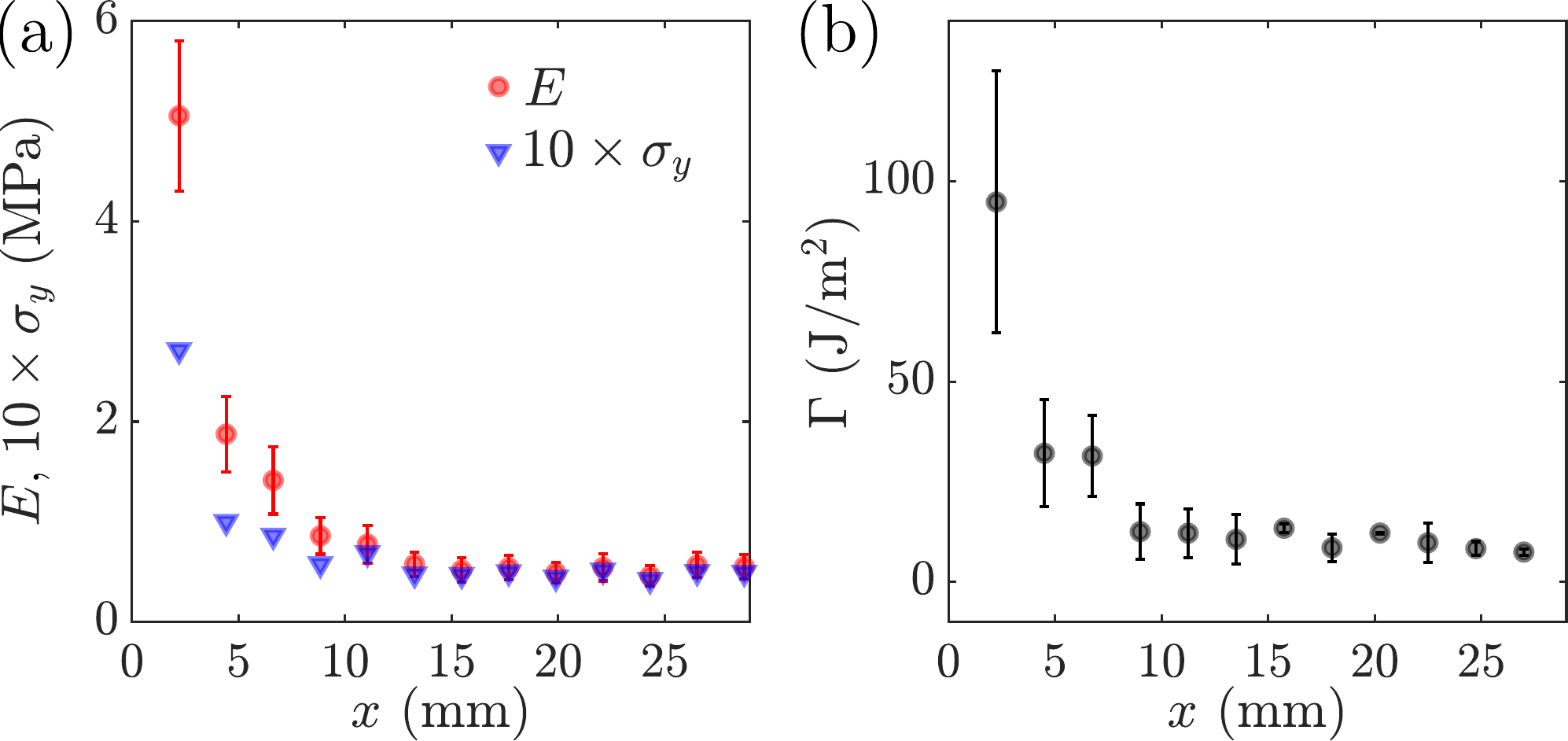}
\caption{\label{fig:E_sigma_Gamma} (a) Measured Young modulus $E$ and yield stress $\sigma_y$; (b) fracture energy as a function of distance $x$ from the periphery.}
\end{figure}

To test this scenario for our cheese, we measured the Young modulus $E$, the yield stress $\sigma_y$ and the fracture energy $\Gamma$ as a function of the distance $x$ from the edge periphery (cf.~Fig.~\ref{fig:E_sigma_Gamma} for results and ESI for measurement protocols). 
The Dugdale length reads $\ell_D = 3.4$ mm in the core and increases up to $6.5$ mm in the edge, so that $h_0 / \ell_D$ is everywhere lower than 0.1.
We thus expect the cutting to systematically be in the plastic shear kinematics. This is indeed confirmed by our real time imaging in Fig.~\ref{fig:merchant_snapshot}a, which clearly shows that both in the core and the periphery the blade touches the crack tip and the chip is sheared up as sketched in Fig.~\ref{fig:sketchTransition}c. We can thus conclude that the observed strong metric variation between the edge and the core of the cheese is not caused by a change from elastic to plastic cutting. But it should rather result from a modulation of the plastic shearing due the radial change of the cheese properties. 

\begin{figure}[ht]
\includegraphics[width=8.5cm]{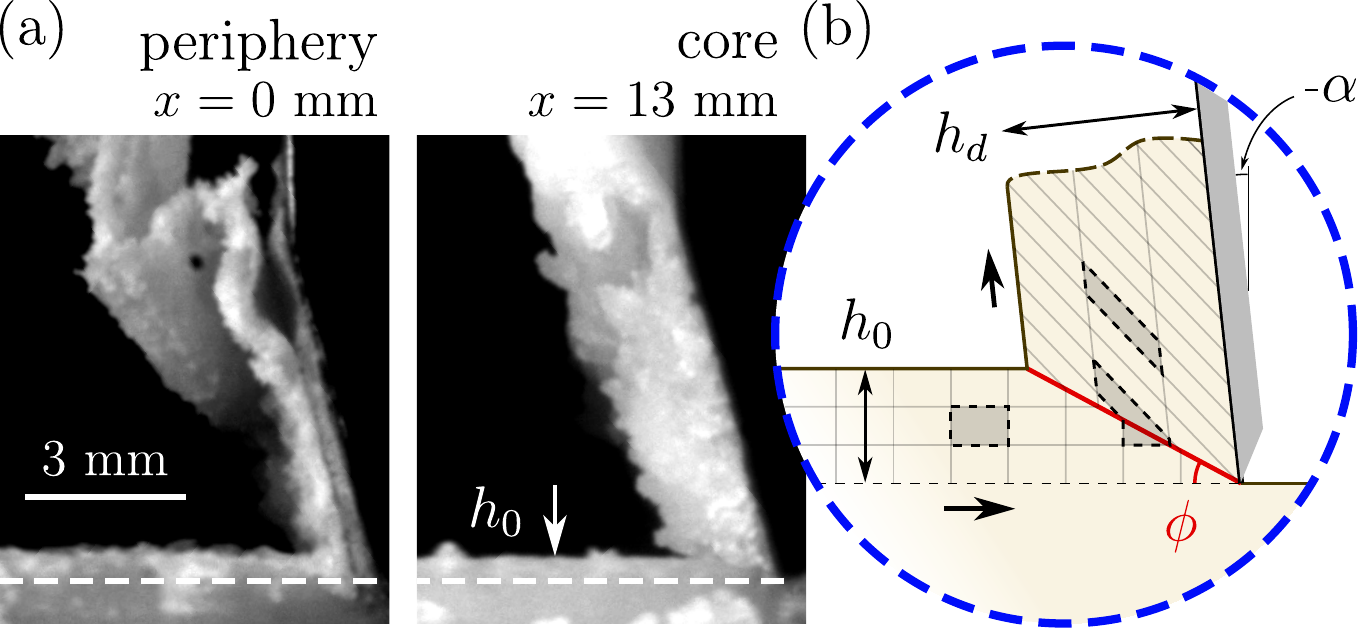}
\caption{\label{fig:merchant_snapshot} (a) Side view of instantaneous snapshots of the cheese layer formation on different radial positions $x$ from the edge periphery, for $\overline{F}_t =0.2$ N/mm. 
For the view at $x=13$ mm in the core region, the outer part of the cheese was removed and the applied weight was adjusted to maintain the same depth of cut $h_0$. The metric ratio $f=\ell_d/\ell_0$ can be directly inferred through $f=h_0/h_d$ under the conservation of volume, with $h_d$ the chip thickness. We remark that the metric is almost conserved in the periphery ($f \sim 1$), while a large increase in thickness is observed in the core ($f \ll 1$), which is consistent with the measured metrics (cf. Fig.~\ref{fig:CR}). 
(b) Sketch illustrating the shear strain in a 2D plastic flow during cutting.
 Three steps in the evolution of a material region labeled in gray: initially rectangular, flowing across the shear plane, and finally after being permanently sheared.}
\end{figure}

-- \textit{Model for the metrics.}  Let us consider the 2D steady-state shear flow 
of a perfectly plastic material (characterized by $\sigma_y$ and with vanishing yield strain, thus neglecting elastic spring back) as shown in Fig.~\ref{fig:merchant_snapshot}b as in several metal cutting models \cite{merchant1941chip,atkins2009science,williams2016fundamentals}.
In the reference frame of the blade, the undeformed material is initially advected towards the blade, and then undergoes an intense shear strain $\gamma_p$ as a result of a change of flow direction along a localized planar zone, named the \textit{shear plane}, tilted at an angle $\phi$. Geometry (together with volume conservation) imposes the value of the plastic shear strain $\gamma_p$ and the metric change $f$ as a function of the shear plane angle $\phi$ and the rake angle $\alpha$ (cf.~Fig.~\ref{fig:gammap} in the End Matter):
%
\begin{gather}
    \gamma_p = \cot(\phi) + \tan (\phi - \alpha)\label{eq:gamma} \\
    f(\phi,\alpha) = {\sin \phi \over \cos (\phi - \alpha)}.\label{eq:f_phi}
\end{gather}

While $\alpha$ is set by the loading fixture, $\phi$ is the only free parameter of the plastic flow and can be determined by minimizing the work provided by the operator on the blade \cite{merchant1941chip, merchant1945mechanics}. From the energy balance for a step advance $dx_c$ of the steady-state cutting process, the work of the operator must equal dissipation (by fracture, plastic flow and friction):
\begin{equation}
F_c dx_c = dU_\Gamma + dU_p + dU_f,
\label{eq:EnBalance}
\end{equation}
we can obtain the expression of the force $F_c$ applied by the operator as a function of the yet unknown shear angle~$\phi$:
\begin{equation}
\begin{array}{r}
F_c(\phi) = b \Gamma + {\sigma_y b h_0 \over 2} [\cot \phi + \tan (\phi - \alpha)] + \\ \\
+(F_c - b \Gamma) {\sin \phi \sin \beta \over \cos(\beta - \alpha) \cos(\phi - \alpha)}.
\end{array}
\end{equation}

The first term $b\Gamma$ is the energy cost for fracture propagation, where $b$ is the transverse dimension of the chip. The second term is the plastic energy associated with the plastic flow along the shear plane at angle $\phi$, where the shear stress is fixed to $\sigma_y/2$ by Tresca yield criterion, and the term in brackets is the plastic shear $\gamma_p$ from Eq.~(\ref{eq:f_phi}). 
The third term is the sliding friction energy between the chip and the blade, where $\beta$ is the friction angle, which is related to Coulomb friction coefficient by $\mu = \tan \beta$ and $F_c-b\Gamma$ is the horizontal component of the force applied by the blade to the plastified chip (cf.~End Matter for full derivations).

We assume that the scrapping process occurs for the minimal operator force (or the minimal dissipation) and $\partial F_c / \partial \phi=0$ leads to the selection of the shear plane angle $\phi_{cut}$~\cite{merchant1941chip}:
\begin{equation}
    \phi_{cut}={\pi \over 4} - {\beta-\alpha \over 2},  \label{eq:Merchant1}
\end{equation}
and finally to the predicted metric:
\begin{equation}
f(\phi_{cut},\alpha)= \frac{\cos (\pi/4 -  (\alpha- \beta)/2)}{\cos (\pi/4- (\alpha+\beta)/2)}
\label{eq:fphicut}
\end{equation}
which surprisingly does not show any dependency on $\sigma_y$ nor on $\Gamma$, which both play a role in the cutting force, but are ruled out in the minimization process. The metrics of the chips only depends on the friction $\beta$ as well as on the rake angle $\alpha$, which is fixed here \cite{williams2010fracture}.

-- \textit{Role of friction on the metrics.} 
It is remarkable that in the absence of friction ($\beta=0$), the shear plane angle reduces to $\phi_{iso} = \pi/4 + \alpha/2$ and lies along the bisector angle between the blade rake and the sample surface. 
This can be associated to the highly symmetric character of the plastic shear flow, as evidenced by the dependency of $\gamma_p$ on the shear plane angle $\phi$ by Eq.~(\ref{eq:gamma}) shown in Fig.~\ref{fig:gammap}. 
The absence of friction would thus systematically lead to isometry ($f=1$), even if the material experiences extremely large plastic shear ($\gamma_p > 2$). 
This is a first proof that friction plays a paramount role, since metric change is the prerequisite for the formation of cheese flowers.
Moreover, the positive nature of $\beta$ (linked to second principle of thermodynamics) breaks the symmetry of the plastic shearing, and implies that in real conditions the shear plane angle will always be smaller than $\phi_{iso}$, leading to metric contraction only ($f\leq 1$ , and $h_d\geq h_0$), as observed in our measurements.
Physically, we see that the chip shortens because  friction during sliding 
along the blade tends to slow the flow. However, a homogeneous metric contraction is not enough to create cheese flowers: according to Eq.~(\ref{eq:fphicut}) a variation of metrics requires a variation of $\beta$.

\begin{figure}[ht]
\centering
\includegraphics[width=7cm]{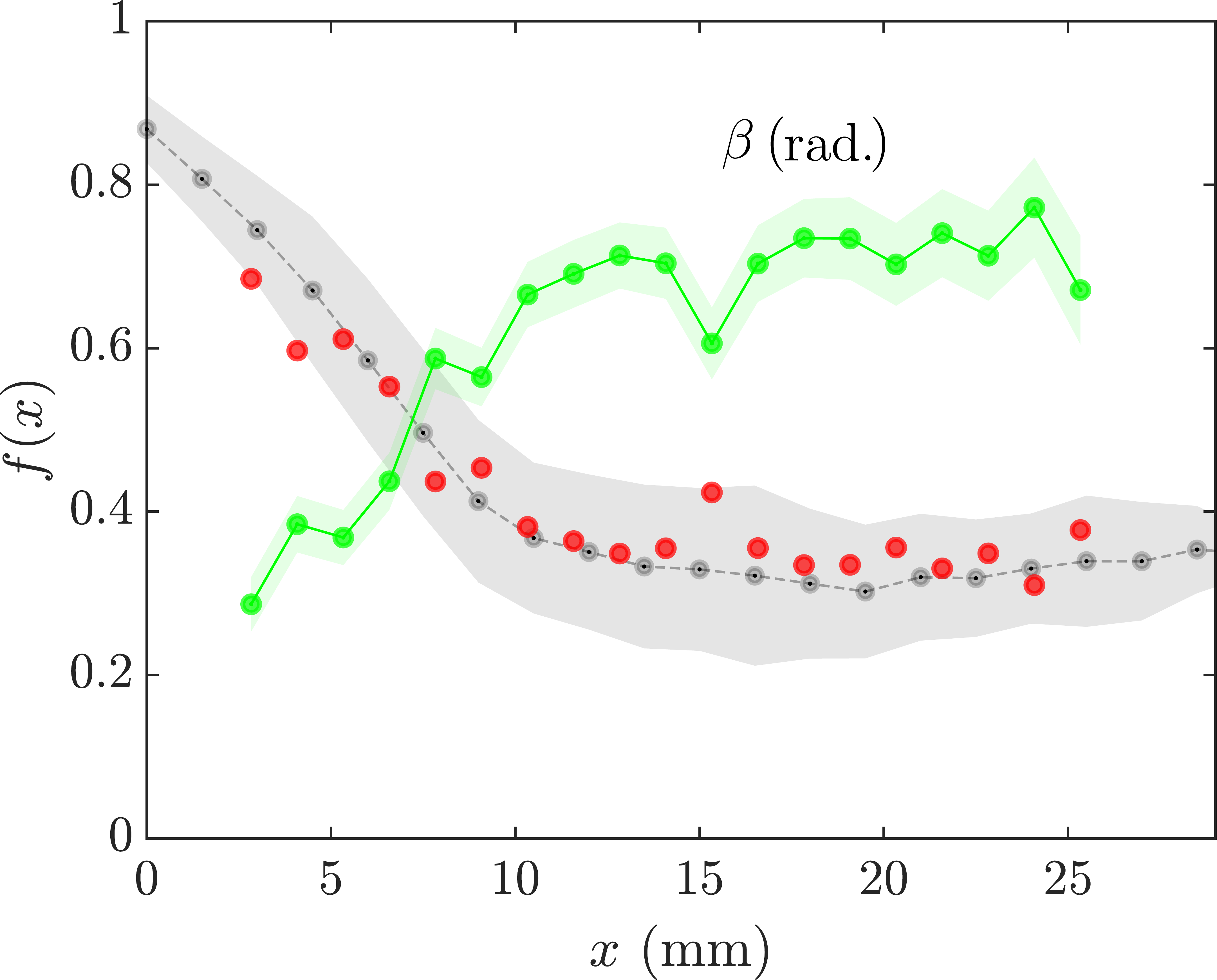}
\caption{\label{fig:CR2} Local measurements of the friction coefficient $\beta$ reported in green, versus  the radial distance $x$ from the edge periphery (the green dots and line represent the local average, and the shaded area the standard deviation). The red circles refer to the prediction of the metrics $f$ by Eq.~(\ref{eq:fphicut}) based on the local measurements of $\beta$. The variation of the measured metric of the cheese flower previously reported in Fig.~\ref{fig:CR}b is shown in gray for comparison.}
\end{figure}

We thus conducted independent measurements of the friction coefficient $\tan \beta$ as a function of the position $x$ on the cheese surface. 
The measurements were performed on the rotating platform with a hex nut under the same conditions for the cheese cutting (details in ESI). 
As reported in Fig.~\ref{fig:CR2}, the friction angle $\beta$ (green) increases from $0.3$ rad to approximately $0.7$ rad within the first 10 millimeter layer and reaches a plateau in the core of the cheese. The estimated metric using Eq.~(\ref{eq:fphicut}) (red) is superimposed on the measured metric of the cheese flower (gray) previously shown in Fig.~\ref{fig:CR}. The agreement is excellent. 
The high value of friction angle in the core of the cheese is responsible for the large permanent orthoradial contraction of the chip $f \sim 0.35$. The strong decrease of friction in the periphery 
is responsible for a gradual reduction of the contraction towards $f \sim 0.9$, which is close to isometry, while remaining fully plastic ($\gamma_p > 2)$. 
The change of metrics that gives origin to the cheese flower should thus be attributed solely to the a gradient of the friction coefficient $\tan\beta$ along the radius of the cheese.  
As a counterproof we performed a new scraping test on the homogeneous soft core region alone, after removing the outer 10 mm of cheese, and we obtained flat thickened chips instead of flowers.

-- \textit{Conclusion.} 
The wrinkly shape of  cheese flowers obtained by scraping the Tête de Moine cheese is due to the irreversible metric change, which is inhomogeneous along the radial direction of the cheese block. 
We attribute the nature of the permanent contraction of the offcuts to the plastic shearing of the chip in the presence of friction along the blade, 
which makes scraping cheese analogous to the continuous ductile cutting of metal chips \cite{trent2000metal,shaw2005metal,stephenson2018metal}. However, flower shaped chips were never reported in metal cutting, since metals have homogeneous properties. 
The inhomogeneous nature of the metric change that leads to the formation of the cheese flowers was shown to be quantitatively determined by the radial variation of the friction coefficient, while being relatively insensitive to the variation of the other mechanical properties such as elastic modulus, yield stress and fracture energy.  
It is however important that the fracture energy is large enough to induce the transition to the plastic shearing regime, which is required for the metric change itself.

The new shaping mechanism evidenced here (inhomogeneous plastic shrinking induced by scraping) is described in its generality. It can be of interest for other materials such as in metal cutting or for polymer materials, when presenting inhomogeneous properties either by formulation, aging or mechanical processing. 
Even for homogeneous materials, the fact that friction properties control the metric change is also particularly interesting for material shaping: starting from a simple homogeneous material, but with a blade designed with spatially varying frictional properties~\cite{Aymard24}, these results open the possibility of programming complex shaping from a simple scraping process.

-- \textit{Acknowledgements.} We thank M. Rabaud for first suggesting this topic, as well as J. Bico, É. Reyssat, F. Lechenault, S. Perrard, A. Eddi, D. Vandembroucq, A. Bouvier, J-W. Hong, T. Gao and M. Saint-Jean for fruitful discussions. This work was partly supported by the Agence Nationale de la Recherche with grants ANR-21-CE33-0018 and ANR-22-CE51-0024.

\bibliographystyle{apsrev4-2}
\newcommand{\noopsort}[1]{} \newcommand{\printfirst}[2]{#1} \newcommand{\singleletter}[1]{#1} \newcommand{\switchargs}[2]{#2#1}

\clearpage
\section{End Matter}
-- \textit{Plastic cutting model.} 
We present here a more complete determination of the plastic cutting model for the orthogonal cutting of a surface layer from a sample made of an homogeneous perfect plastic material (characterized by a yield stress $\sigma_y$ and infinite elastic modulus) and fracture toughness $\Gamma$, inspired by \cite{merchant1941chip,merchant1945mechanics,atkins2009science,williams2010fracture}. 
The sample moves with a horizontal velocity $\overrightarrow{V_c}$ against a stiff blade, which has an almost vertical rake angle $\alpha$ as sketched in Fig.~\ref{fig:schemaMerchant}. The vertical position of the cutting edge is fixed at a depth of cut~$h_0$ with respect to the sample surface. As sketched in Fig.~\ref{fig:merchant_snapshot}b, the kinematics of the plastic flow is modeled as a homogeneous plastic shear of the chip that is considered to instantly happen when the material crosses a thin localized shear plane, which is tilted at an angle $\phi$ 
\cite{merchant1945mechanics}.
After crossing the shear plane, the plasticized chip is moving parallel to the rake face with a spatially homogeneous velocity $\overrightarrow{V}_{chip}$ at an angle $\alpha$. When crossing the shear plane, the chip undergoes a discontinuity in the velocity $\overrightarrow{V_s} = \overrightarrow{V}_{chip}-\overrightarrow{V_{c}}$, which is directed along the angle $\phi$ (cf.~Fig.~\ref{fig:schemaMerchant}a). 

\begin{figure}[ht]
\includegraphics[width=8.5cm]{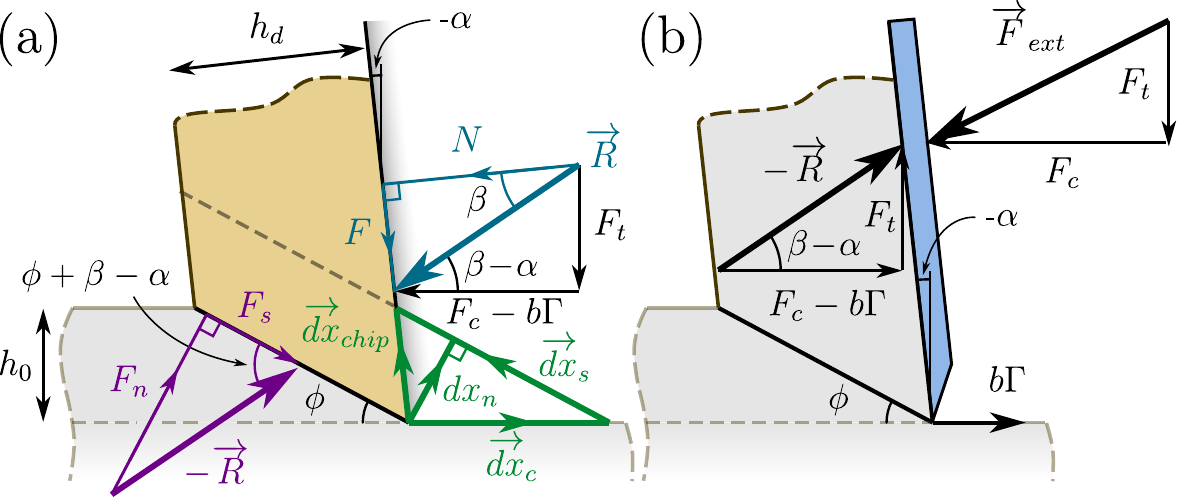}
 \caption{\label{fig:schemaMerchant} Free body diagrams for the force balance on the chip (a) and the blade (b). The energy for fracture propagation is associated with a horizontal force $b \Gamma$ acting on the blade tip. The green triangle in (a) represents the incremental displacements $\protect\overrightarrow{dx}_i = \protect\overrightarrow{V_i}dt$ of the cheese, chip and shear plane.}
\end{figure}

At this stage of the modeling, the value of the shear plane angle $\phi$ is undetermined and will be derived later. 
By enforcing volume conservation of the chip, and applying some geometric relations, we can deduce that the thickness $h_d$ of the chip will be modified into:
\begin{equation}
    h_d = h_0 {\cos (\phi - \alpha) \over \sin \phi},  \label{eq:EM_hd}
\end{equation}
and the length of the chip will change by the ratio:
\begin{equation}
    f(\phi,\alpha) = {\ell_d \over \ell_0} = {h_0 \over h_d} = {\sin \phi \over \cos (\phi - \alpha)},  \label{eq:EM_fphi}
\end{equation}
which defines the metric of the plastified chip. 

As a consequence the sliding velocity of the chip along the blade is also altered by the same factor: 
\begin{equation}
    V_{chip} = f(\phi,\alpha) V_{c}.  \label{eq:EM_vchip}
\end{equation}

The homogeneous plastic shear strain $\gamma_p$ of the chip can be obtained by evaluating the ratio between the relative displacement $dx_s$ of the plastified chip parallel to the slip plane $\phi$ and the distance $dx_n$ from the slip plane, for a displacement $dx_c$ of the cheese. The displacement vectors $\overrightarrow{dx}_{c}$,   $\overrightarrow{dx}_{chip}$ and $\overrightarrow{dx}_{s} = \overrightarrow{dx}_{chip} - \overrightarrow{dx}_{c}$ can be estimated geometrically from the green triangle in Fig. \ref{fig:schemaMerchant}a. Their amplitudes read:  
\begin{eqnarray}
    dx_s = dx_{c} \cos \phi + dx_{chip}\sin (\phi - \alpha),\\
    \label{eq:EM_dxs}
    dx_{n} = dx_{chip} \cos(\phi - \alpha) = dx_{c} \sin \phi,\\
    \label{eq:EM_dxchip}
    \gamma_p = {dx_{s} \over dx_{n}} =\cot \phi + \tan (\phi - \alpha).
    \label{eq:EM_gamma}
\end{eqnarray}

\begin{figure}[ht]
\includegraphics[width=7cm]{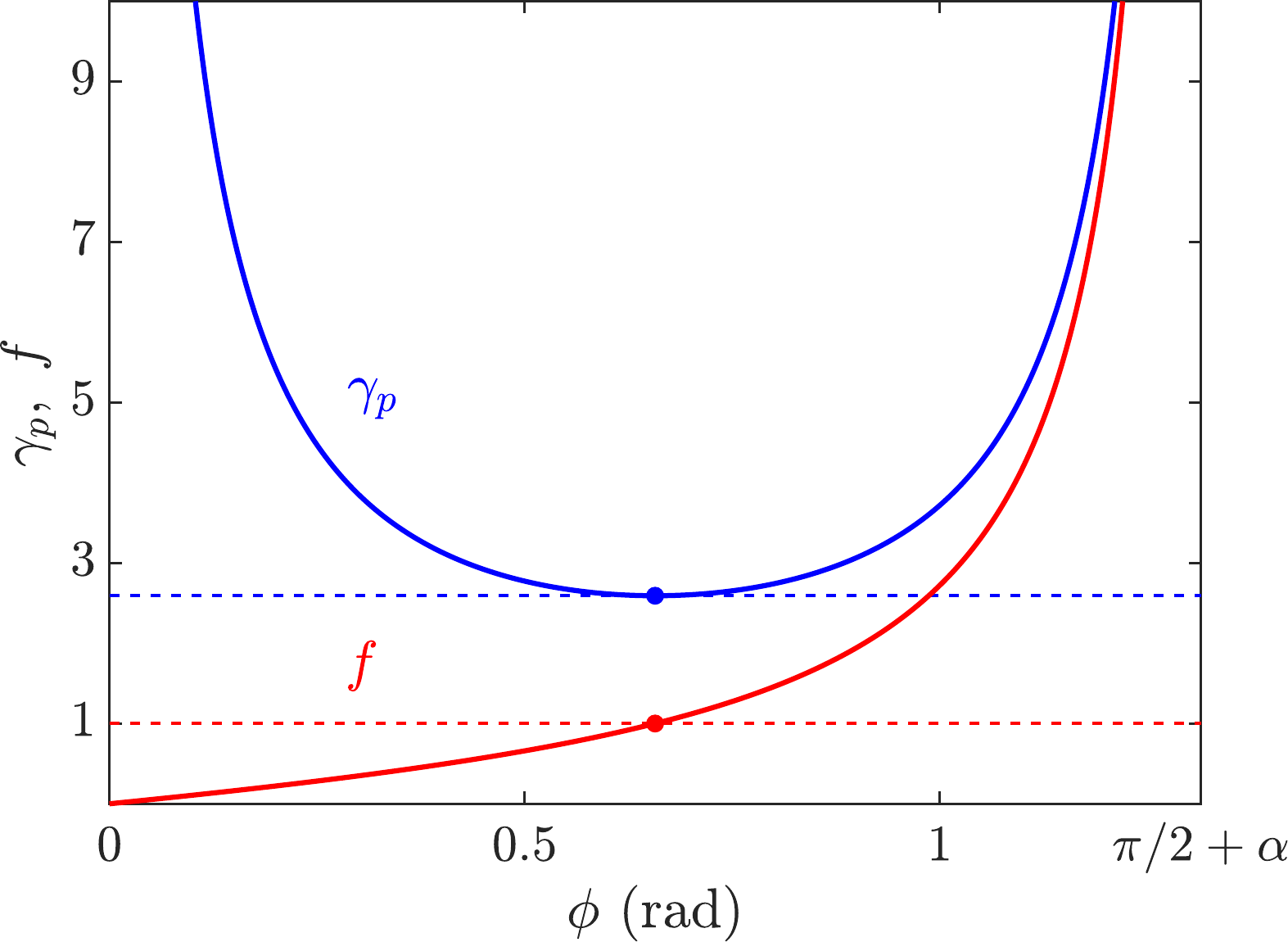}
\caption{\label{fig:gammap} The variations of the plastic shear strain $\gamma_p$ (blue) and the metric $f$ (red) as a function of the shear plane angle $\phi$ from Eqs.~\ref{eq:gamma} and \ref{eq:f_phi}. As $\phi$ reaches the plane of symmetry between the cheese's upper surface and the blade $\phi=\pi/4+\alpha/2$, $\gamma_p$ reaches its minimum value of approximately $2.6$ whereas the metric reaches 1.}
\end{figure}

We remark that according to the plastic shear kinematics, all morphogenetic parameters $h_d$, $f$ and $\gamma_p$ are determined from the knowledge of $\alpha$ and $\phi$ as plotted in Fig.~\ref{fig:gammap}. While the rake angle $\alpha$ is a fixed loading parameter, the shear plane angle $\phi$ is the only free variable and its determination requires  mechanical arguments.  

When the tool moves a distance $dx_{c} = V_{c} dt$ so does the cut, and we may define the energy balance:
\begin{equation}
dW = dU_\Gamma + dU_p + dU_f,
    \label{eq:EM_dW}
\end{equation}
where $dW$ is the work of the external applied force,   $dU_\Gamma$ is the energy for creating new fracture surfaces, $dU_p$ is the energy dissipated in plastic deformation of the chip and $dU_f$ is the energy dissipated in frictional sliding between the chip and the blade:
\begin{equation}
\left\{
\begin{array}{l}
  dW = \overrightarrow{F}_{ext} \cdot \overrightarrow{dx}_{c} = F_{c} \cdot dx_{c} \\
  dU_\Gamma = b \Gamma \cdot dx_{c} \\
  dU_{p} = \overrightarrow{R} \cdot \overrightarrow{dx}_{s} = F_s \cdot dx_s \\
  dU_{f} = -\overrightarrow{R} \cdot \overrightarrow{dx}_{chip} = F \cdot 
 dx_{chip} 
\end{array}
\right.
    \label{eq:EM_works}
\end{equation}

As illustrated by the free body diagrams in Fig.~\ref{fig:schemaMerchant}, the external cutting force $\overrightarrow{F}_{ext}$ has horizontal component $F_{c}$ and transverse component $F_t$, the latter does not contribute to the work. 
In the plastic shear regime, the tip of the blade is in contact with the fracture point (which is referred to as the {\it crack tip touching condition}). The fracture energy can thus be associated with the work of a horizontal force $b\Gamma$ acting on the blade tip \cite{williams2010fracture}, where $b$ is the transverse dimension of the chip. The net horizontal component of the force $\overrightarrow{R}$ transmitted from the blade to the plastified chip is thus $F_{c}-b\Gamma$. This can be decomposed into a normal force $N$ and a tangent friction force $F = N \tan \beta$, where $\mu =\tan \beta$. By equilibrium, the reacting force applied by the substrate on the plastified chip though the shear plane is $-\overrightarrow{R}$, and it can be decomposed into a tangent shear stress $F_s$ and a normal force $F_n$.




The shear stress on the plastic shear plane is fixed to $\sigma_y/2$ according to the Tresca yield criterion. Since the area of the shear plane is $A_s = b h_0/\sin \phi$, the force along the shear plane is: 
\begin{equation}
F_s = 
{\sigma_y \over 2} {bh_0 \over \sin \phi},
\label{eq:EM_Fs}
\end{equation}
and the plastic shear energy is (using Eqs.~(\ref{eq:EM_dxs}), (\ref{eq:EM_works}) and (\ref{eq:EM_Fs})):
\begin{equation}
dU_{p} = {\sigma_y bh_0 \over 2} [\cot \phi + \tan (\phi - \alpha)]dx_{c}.
\label{eq:EM_dUplast}
\end{equation}
%



When using Coulomb friction law with $\mu = \tan \beta$, the friction force $F$ can be written as: 
\begin{equation}
F = R \sin \beta = {F_c - b \Gamma \over \cos(\beta - \alpha)} \sin \beta. 
\label{eq:EM_Ffrict}
\end{equation}

The energy dissipated in frictional sliding between the chip and the blade is thus (using Eqs.~(\ref{eq:EM_vchip}), (\ref{eq:EM_works}) and (\ref{eq:EM_Ffrict})):
\begin{equation}
dU_{f} = (F_c - b \Gamma) {  \sin \beta \sin \phi \over \cos(\beta - \alpha) \cos(\phi - \alpha)} dx_{c}.
\label{eq:EM_dUfr}
\end{equation}

Combining these into the energy balance given in Eq.~(\ref{eq:EM_dW}): 
\begin{equation}
\begin{array}{r}
F_c = b \Gamma + {\sigma_y b h_0 \over 2} [\cot \phi + \tan (\phi - \alpha)] + \\ \\
+(F_c - b \Gamma) {\sin \phi \sin \beta \over \cos(\beta - \alpha) \cos(\phi - \alpha)},
\end{array}
\end{equation}
%
%
\begin{equation}
F_c = b \Gamma + {\sigma_y b h \over 2} {\gamma_p(\phi,\alpha) \over Q(\phi,\alpha,\beta)},
\label{eq:EM_Fcutphi}
\end{equation}
where
\begin{equation}
Q(\phi,\alpha,\beta) \equiv 1 - {\sin \phi \sin \beta \over \cos(\beta - \alpha) \cos(\phi - \alpha)}.
\label{eq:EM_Q}
\end{equation}

In classical cutting models \cite{merchant1941chip,merchant1945mechanics,atkins2009science,williams2010fracture} both the rake angle $\alpha$ and the depth of cut $h_0$ are fixed as in a real orthogonal cutting machine. The friction angle $\beta$ is fixed and depends on the specific material to be cut and on the blade surface finish. 
The slip angle $\phi$ is the only free variable and it is determined by minimizing the power of the cutting process $
P = F_c V_c$ \cite{merchant1941chip}, which is equivalent to minimize the horizontal cutting force $F_c$, since the cutting velocity $V_c$ is a constant loading parameter. Considering Eq.~(\ref{eq:EM_Fcutphi}), all the dimensional quantities disappear in the minimization process: 
%
\begin{equation}
{d \over d\phi} {\gamma_p(\phi,\alpha) \over Q(\phi,\alpha,\beta)} = 0,
\label{eq:EM_minimize}
\end{equation}
and the value of the shear plane angle $\phi_{cut}$ is only dependent on the dimensionless values of $\alpha$ and $\beta$:
\begin{equation}
\phi_{cut} = {\pi \over 4} - {\beta - \alpha \over 2}.
\label{eq:EM_phicut}
\end{equation}

\

By substituting into Eq.~(\ref{eq:EM_Fcutphi}) we obtain the prediction for the cutting force: 
\begin{equation}
F_c = b \Gamma + {\sigma_y b h_0} {\cos (\beta - \alpha) \over 1 + \sin (\alpha - \beta)},
\label{eq:EM_CutForce}
\end{equation}
as well as the metric change: 
\begin{equation}
    f(\phi_{cut},\alpha) = 
    \frac{\cos (\pi/4 -  (\alpha- \beta)/2)}{\cos (\pi/4- (\alpha+\beta)/2)},
\label{eq:EM_f_phicut}
\end{equation}
which turns out to be independent of the depth of cut $h_0$, as well as on most cheese properties except the cheese/blade friction coefficient $\tan \beta$.  

While in orthogonal cutting machines the depth of cut $h_0$ is fixed and the transverse force is proportional to $h_0$:
\begin{equation}
F_t = (F_c - b \Gamma) \tan(\beta-\alpha) = {\sigma_y b h_0} {\sin (\beta - \alpha) \over 1 + \sin (\alpha - \beta)}
\label{eq:EM_Ft}
\end{equation}
in our cheese cutting setup, the transverse force is set by the applied weight, and the depth of cut increases until reaching the equilibrium with the upward drag of the friction force on the blade, in order to respect the same Eq.~(\ref{eq:EM_Ft}).


\end{document}


\title{Morphogenesis of cheese flowers through scraping- Electronic Supplementary Material}
\author{J.~Zhang}
\email{jishen.zhang@espci.fr}
\affiliation{PMMH, CNRS, ESPCI Paris, Université PSL, Sorbonne Université, Université de Paris, F-75005, Paris, France}
\author{A.~Ibarra}
\affiliation{PMMH, CNRS, ESPCI Paris, Université PSL, Sorbonne Université, Université de Paris, F-75005, Paris, France}
\author{B.~Roman}
\affiliation{PMMH, CNRS, ESPCI Paris, Université PSL, Sorbonne Université, Université de Paris, F-75005, Paris, France}
\author{M.~Ciccotti}
\email{matteo.ciccotti@espci.fr}
\affiliation{SIMM, CNRS, ESPCI Paris, Université PSL, Sorbonne Université, F-75005, Paris, France}
\date{\today}

\maketitle

\begin{figure}[ht]
\includegraphics[width=8cm]{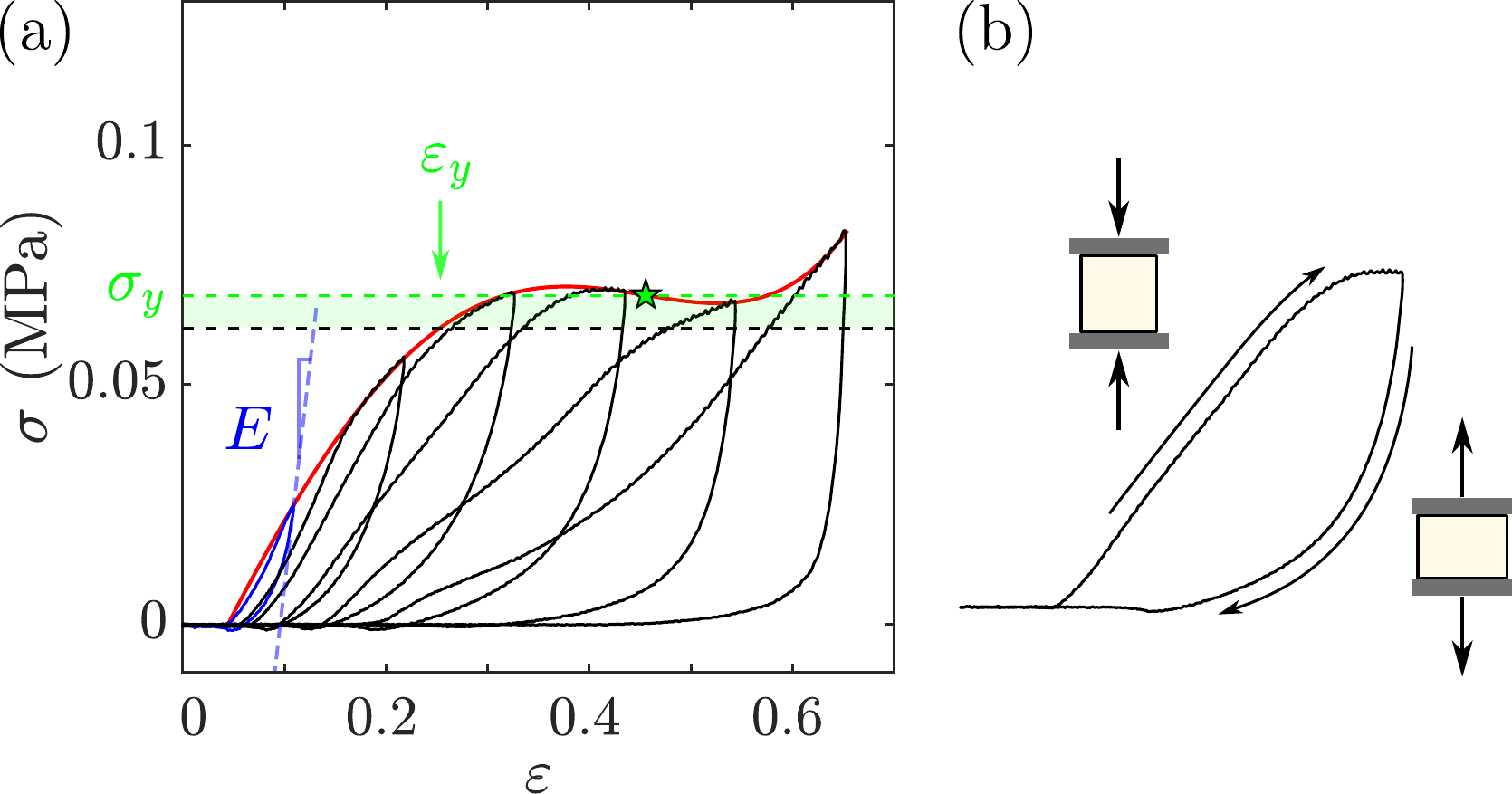}
\caption{\label{fig:multicycle} (a) Example of elasto-plastic properties measurements by multi-cycle compression test with a typical loading-unloading cycle illustrated in (b).}
\end{figure}
\begin{figure}[ht]
\includegraphics[width=8.5cm]{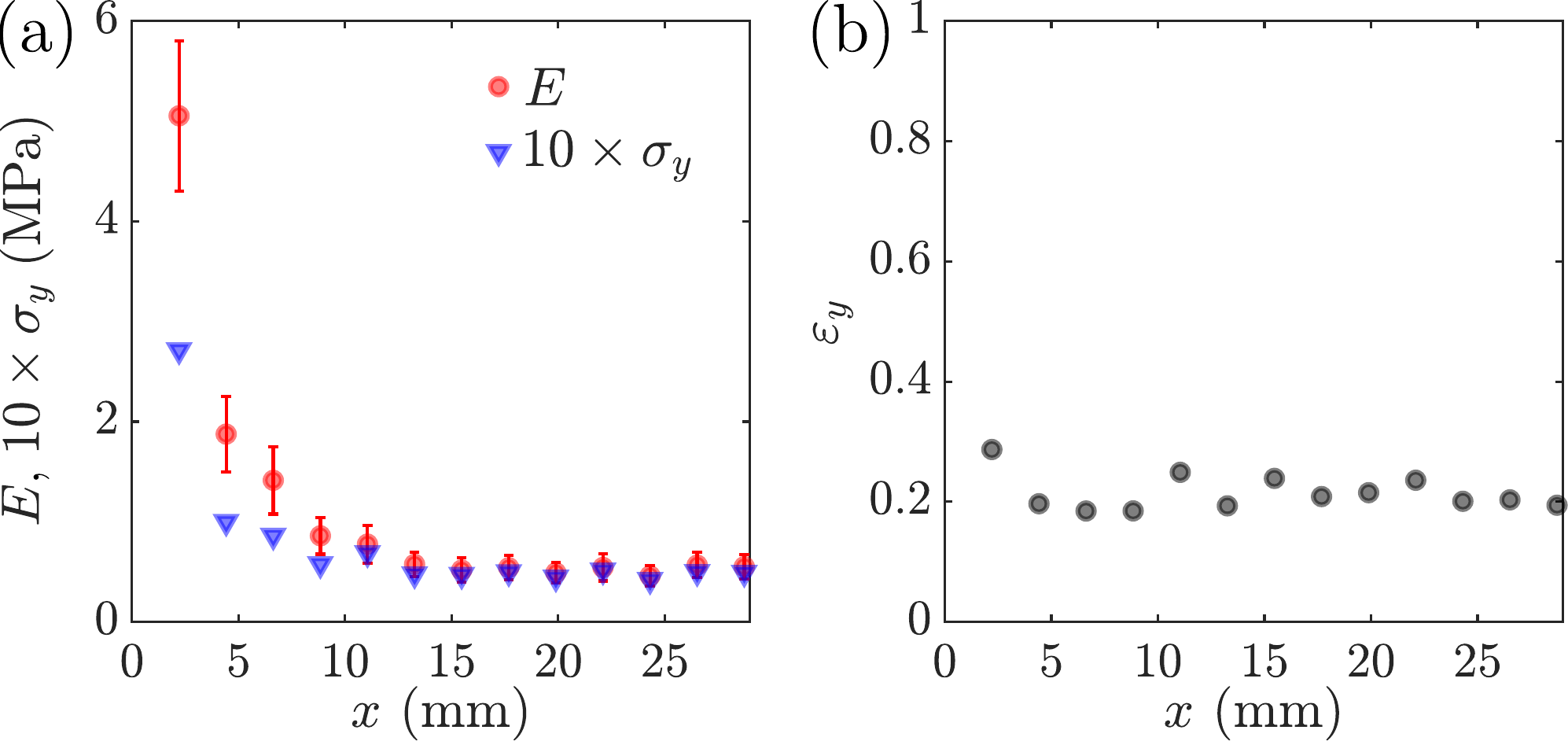}
\caption{\label{fig:yield} Young modulus $E$, yield stress $\sigma_y$ (a) and yield strain (b) as a function of distance $x$ from the periphery.}
\end{figure}

-- \textit{Elasto-plastic behavior.}
In order to characterize the elastic and plastic behavior of the Tete de Moine's cheese we performed cyclic uni-axial 
compression tests on prismatic cheese samples ($4.6 \times 4.6 \times 4.4$ mm$^3$) 
extracted at different distances $x$ from the periphery. Several cycles at increasing maximum strain were applied with a constant loading speed of 0.05 mm/s. Measurements from a typical test on a cheese sample at $x=11$ mm from the edge periphery are shown in Fig.~\ref{fig:multicycle}. 
Although the cyclic response is complex and presents some hysteresis even at low applied strain, we can clearly observe the establishment of a well defined plastic plateau for strains larger than 0.2 and up to 0.6. We did not push the investigation to larger strains, since the friction and damage on the sample made the data non trustable. However, since our main aim is to test the validity of the plastic cutting model, based on large strain shearing, it is enough to derive reasonable values of $E$ and $\sigma_y$ as a function of the position $x$. 
Since the plastic plateau is well defined, we estimated $\sigma_y$ by the inflexion point (green star) from a polynomial envelope fit (red curve) on the maximum stress-strain of each cycle. The Young modulus $E$ was estimated by a linear fit on the unloading part for strain around 0.1 (blue curve).
The yield strain values $\varepsilon_y$ were estimated by intersecting the envelope stress-strain curve with the horizontal black dashed line at $\sigma = 0.9\sigma_y$. 
The values of Young modulus $E$, yield stress $\sigma_y$ and yield strain $\varepsilon_y$ are reported in Fig.~\ref{fig:yield} as a function of the radial distance $x$ from periphery.

Both measurements reveal homogeneous properties after a boundary layer of about 10 mm, with average plateau values of $E = 560 \pm 85$ kPa and $\sigma_y = 49 \pm 8$ kPa. Both $E$ and $\sigma_y$ increase significantly close to the external periphery (by more than a factor of 5), however their ratio, giving the yield strain $\varepsilon_y \sim \sigma_y/E$,   is quite constant all along the position $x$ with a value of $0.21 \pm 0.03$. We remark that this value is very small in front of the large plastic strains  that are experienced during the plastic scraping ($\gamma_p>2$, cf.~Fig.~7 in the main text). This justifies neglecting the elastic springback in the modeling. 

\begin{figure}[ht]
\includegraphics[width=8.5cm]{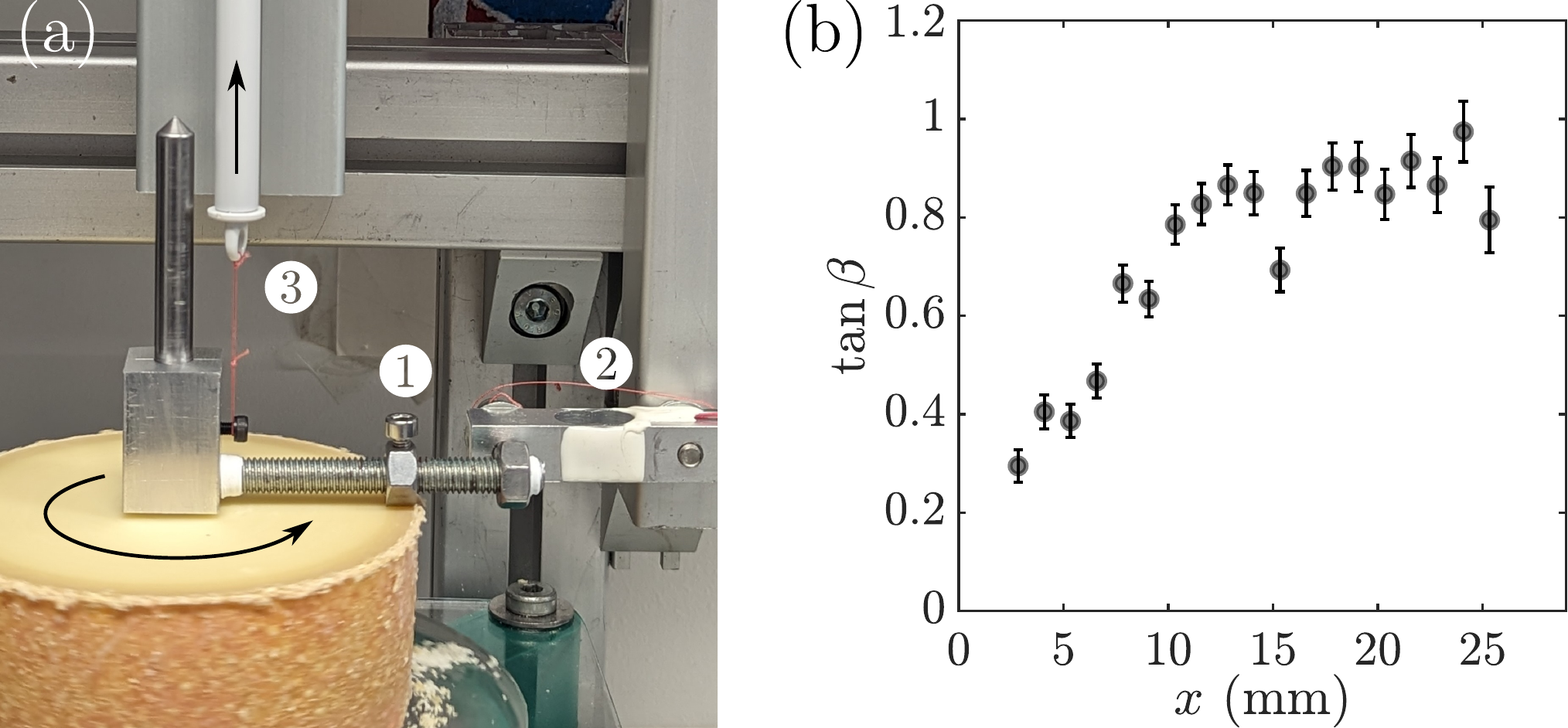}
\caption{\label{fig:beta} (a) Configuration of the custom cutting setup for measuring local friction: the blade is substituted with a hex bolt  (1), the horizontal force opposing to friction is measured by a beam sensor (2), and upward lifting spring force gauge (3) is used to set the normal weight. (b) Measurements of the local friction coefficient $\tan\beta$ as a function of distance $x$ from the periphery under a normal load of 3 kPa.}
\end{figure}

-- \textit{Friction coefficient.}
Local measurements of the Coulomb friction coefficient $\mu = \tan\beta$ were performed by using the same custom setup as for the main experiment and replacing the blade by a screw thread on which a hex nut, used as the slider, was set at a distance $x$ from the periphery, which could be changed by rotating the bolt (cf. Fig.~\ref{fig:beta}a). 
The surface of the bottom facet in contact with the cheese reads 17 $\mathrm{mm^2}$. 
In order to avoid indentation of the slider in the cheese, we were limited to a maximum applied weight corresponding to a normal stress of 3 kPa, which is lower than the normal stress applied on the blade during the cutting experiments. This low value was achieved by using an upward lifting spring force gauge (cf. Fig.~\ref{fig:beta}a) to reduce the effective weight of the slider fixture. 
A beam-type load cell was installed to measure the horizontal force received from the thread's end during the sliding test.  
To remove the residual torque due to the friction on the central rod, a no-load test was first conducted using the spring suspension to detach the slider fixture from the cheese surface.

The local friction coefficient $\tan \beta(x)$ was estimated as the ratio between the measured net horizontal force on the slider and the effective weight. The result is reported in Fig.~3b as a function of the distance $x$ from the periphery. Interestingly, we observe an increasing trend of $\tan\beta$ from the periphery to the core, with a variation of a factor 3. 

\begin{figure}[ht]
\includegraphics[width=8.5cm]{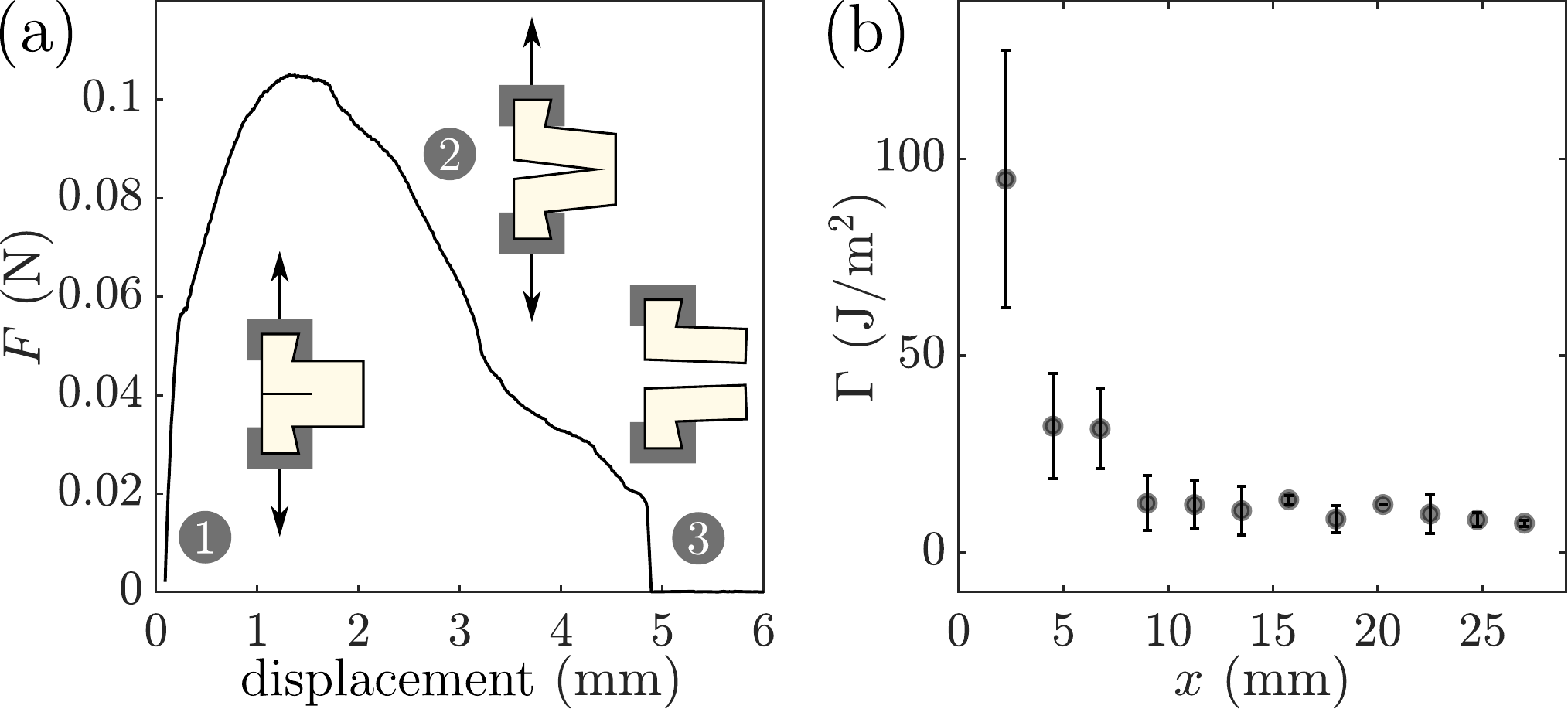}
\caption{\label{fig:fracture_h} (a) Example of force/displacement curve during the fracture testing of a 
cheese sample. (b) Measured fracture energy as a function of the distance $x$ from the cheese edge.}
\end{figure}

-- \textit{Fracture energy.} To estimate the local fracture energy of the cheese material, we performed crack opening tests on cheese samples extracted at  different distances $x$ from the periphery. The sample shape is close to the compact tension geometry with two side wedges designed to fit the grips (cf.~Fig.~\ref{fig:fracture_h}a). The uniform thickness of the samples is $t=4.5$ mm, the width is $w =4$ mm the length is 8 mm and an initial flaw of 4 mm is cut into the sample. We applied a traction displacement perpendicular to the crack with a constant loading speed of 0.1 mm/s. An example of the measured force/distance curve is shown in Fig.~\ref{fig:fracture_h}a for a cheese sample extracted at 9 mm from the edge periphery, covering the crack opening (1), propagation (2) and a total split (3) of the cheese sample. 

At the end of each test the two separated cheese parts can be recombined to fit into the original sample shape, as in brittle fracture, we can thus neglect large scale yielding of the sample and use Linear Elastic Fracture Mechanics to process the data. Since the crack propagates in a stable and quasi-static manner up to the end of the sample where the load vanishes, the fracture energy $\Gamma$ can be approximated  by integrating the total work on the force/distance curve and dividing by the area of the newly created fracture surface $\Gamma = 1/lt \int_{0}^{x_{max}} F(x) \,dx$, where $l$ is the initial length of the intact ligament. The measured values of $\Gamma$ are shown in Fig.~\ref{fig:fracture_h}b. In the same manner as for $E$, $\sigma_y$ and $\beta$, we can distinguish a plateau value of $\Gamma = 10 \pm 2~\mathrm{J/m^2}$ in the core of the cheese and an increase in the cheese boundary layer of 10 mm from the periphery, up to a maximum of $\Gamma = 90 \pm 30~\mathrm{J/m^2}$.
We conclude that all these cheese properties are homogeneous in the core of the cheese and altered in the first 10 millimeter layer close to periphery as a result of the drying process.